# On Arrival: Challenges and Opportunities Around Early-Stage Resettlement of Refugees in Australia

PINYAO SONG, University of Queensland, Australia
APARNA HEBBANI, University of Queensland, Australia
DHAVAL VYAS, University of Queensland, Australia

When refugees arrive in a host country, the form of immediate help and support they receive from various service providers sets the stage for successful settlement, integration, and social cohesion. This paper presents results from an exploratory study that investigated refugees' perceptions of initial services received upon migration – in the first six months of their arrival. In collaboration with a refugee settlement services provider, we engaged 12 newly-arrived refugees in a qualitative study that employed a photo-diary study and semi-structured interviews. Based on our findings, we present refugees' experiences over three phases: immediate services upon arrival (first day); initial-settlement experiences (first week); and ongoing settlement experiences (up to twelve months). Through an in-depth unpacking of these phases, we show ongoing efforts and challenges associated with resettlement; and present implications for design for CSCW researchers.

## 1 INTRODUCTION

Australia is a popular destination for migrants. The latest Australian Bureau of Statistics (ABS) census data indicates that in the period extending from 2000 to 2022, Australia welcomed approximately 3 million permanent migrants, with 9% of this comprising of humanitarian migrants [9]. Within Australia's national context, a "humanitarian entrant" refers to individuals who enter the country through its Humanitarian Programme, which encompasses refugees and other individuals with humanitarian needs [10]. While COVID-19 halted most international arrivals into Australia, recent conflicts in Ukraine, Syria, Ethiopia, and Afghanistan, resulted in Australia re-opening its doors to people fleeing these countries and seeking a refugee status. According to the United Nations High Commission for Refugees (UNHCR), as of mid-2022, the total number of individuals subjected to forced displacement surged to 103 million [58]. Governmental or non-governmental organisations (NGOs) have been making efforts to assist refugees' early-stage resettlement. Australia has two government programs that support newly-arrived refugees: the Special Humanitarian Program (HSP) and the Settlement Engagement and Transition Program (SETS). The HSP encompasses initial services to help refugees begin their lives in Australia, such as airport pickup, immediate accommodation assistance, assistance in finding permanent housing, and connecting refugees with essential public resources. These resources include Australia's social security agency, Centrelink – an Australian Government department that delivers economic support and other payments to Australians, banks, public transport options, translating and interpreting services, healthcare services, educational opportunities, employment assistance, and childcare services [10]. The SETS program provides support for post-arrival refugees who have come to Australia alone. It consists of two parts: client services, which offer refugees settlement-related information, advice, advocacy, and assistance in accessing mainstream and other relevant services; and community capacity-building services, which aim to enhance refugees' social participation, economic prospects, and overall personal well-being, ensuring long-term positive settlement outcomes [59]. In other countries such as the U.S., the initial services provided within 30 days upon arrival include basic food, clothing, shelter, orientation and referral to social service agencies [26]. Ongoing services, encompassing English language training, case management, and employment support, are accessible during the initial 3 to 12 months and can be extended up to 5 years for refugees who are unemployed or receiving welfare assistance [35]. Thus, non-governmental service providers and organisations in the multicultural space, play a crucial role in supporting refugees, by providing mediating support to connect refugees with essential public resources.

Current research in Computer Supported Cooperative Work (CSCW) and Human-Computer Interaction (HCI) has primarily focused on understanding and addressing the multifaceted challenges faced by refugees through the development of technological solutions. These challenges encompass language barriers, cultural disparities, limited access to essential resources and services, employment discrimination, and housing difficulties during the initial resettlement phase [60]. Information and Communication Technologies (ICTs) have played an important role in aiding refugees [13, 52], and emerging technologies like extended reality are proposed to have great potential in assisting them [1, 40]. While HCI research has paid significant attention to the services themselves and their accessibility, most studies tend to be narrowly focused on a specific service or challenge, such as healthcare, information-seeking, or interpreting services [13, 23, 53]. There is limited coverage of the understanding of how newly-arrived refugees perceive the support provided by settlement agencies.

Therefore, within this context, we aimed to understand the experiences of newly-arrived refugees with the services provided by a mediating service provider, understanding their perspectives of these services to identify potential areas for improvement and explore preliminary design opportunities. To achieve this, we recruited 12 participants who had been in Australia for 6 months, by collaborating with a refugee settlement service provider. The overarching research question that we aim to explore in this paper is 'how do newly-arrived humanitarian entrants perceive the settlement services provided by the service provider in Australia?'. Data were gathered by employing a two-stage qualitative approach that combined a photo-diary exercise, where participants were asked to take photos to record their experiences, followed by semi-structured interviews to further unpack/understand these experiences. Our data analysis is presented across three distinct phases of service provision: immediate services upon arrival, initial services, and ongoing services. The immediate services upon arrival include our participants' experiences on the first day of their arrival in Australia (e.g., hotel accommodation, giving a basic essential pack, and induction of the local area for buying groceries and using public transportation). The initial services comprise connecting refugees to essential public resources, such as registering with local municipalities, medical insurance, and welfare agencies, and enrolling them into training programs around education, employment, and language learning. The ongoing services include providing long-term accommodation and exploring employment opportunities to enable settlement in the host country.

This paper makes two contributions to the current CSCW & HCI research on the refugee discourse: 1) It provides an empirical account of newly-arrived refugees' experience in accessing services provided by non-governmental organizations. There are limited studies in the HCI domain where the mediating service providers' role is discussed in relation to refugee research. 2) It identifies areas where services offered by service providers can be enhanced and proposes design implications to address these problems.

## 2 BACKGROUND

### 2.1 Refugees in HCI & CSCW

Current research in CSCW and HCI that focuses on refugee resettlement largely aims to address various challenges associated with the process [45]. These challenges range from mistrust and skepticism from local communities to cultural barriers, trauma from displacement, and limited social resources [5, 6, 45]. For example, Talhouk et al. investigated the experiences of refugees in Lebanon using the electronic voucher system for food aid and highlighted the existing issues of the system such as power and information asymmetry [51]. Weibert et al. proposed the Arrival Literacy concept, referring to mastering the necessary skills to navigate various forms of text (digital and paper-based) for dealing with bureaucratic and administrative tasks, and argued such skills are crucial for establishing a sense of belonging [56]. Fisher et al. summarized the principles of the humanitarian research framework, after years of engagement in conflict zones, and pointed out the issue that the young generation suffered from disrupted education leading to low literacy levels [21]. To aid refugees in overcoming these challenges, researchers have developed and explored technologies primarily for educational and skill-building purposes. For example, certain studies have promoted cultural engagement and preservation by making [32], and employing 3D digital fabrication techniques to enhance children's education [47], and focused on maintaining cultural identities, health, and well-being [22, 45]. Lee et al. studied how RAS (refugees and asylum seekers) in Australia develop their entrepreneurship and suggested that technologies have the potential for upskilling refugees, and benefitting their entrepreneurship [31]. Moreover, researchers have delved into the difficulties refugees face in adopting technologies in host countries. One notable study by Sabie et al. used three years of fieldwork in Canada to outline the challenges refugees experience when accessing essential services through digital platforms [42]. HCI research has also ventured into fostering empathy and public engagement. Studies like those by Norman et al. used CityScope, a platform for urban planning, to facilitate public participation in decisions related to refugee accommodation [38]. Additionally, Kors et al. utilized a Mixed-Reality game to spotlight the struggles that refugees endure, suggesting that Extended Reality (XR) could serve as a powerful medium for generating empathy [29]. Virtual Reality (VR) has been particularly effective as an empathy-building and knowledge-building tool. The 360-degree film "Clouds Over Sidra" allows viewers to immerse themselves in the experiences of residents of the Zaatari Refugee Camp, which houses 130,000 Syrian refugees [61]. In addition, Virtro Technology [62] has developed VR programs to assist refugees and immigrants with language acquisition and job readiness. The use of mobile technologies is also beneficial for refugees' resettlement process, including promoting social integration and providing chances to improve their access to valuable information that can support their daily routines [3]. For example, Talhouk et al. pointed out that refugees have varying informational needs at different times in their adaptation journeys, and technologies should facilitate the sharing of both general and local knowledge [53]. ICTs have been widely deployed to support refugees [30], such as helping refugees find essential information [7, 23], treating mental stress problems [1, 19, 48], building social networks [5, 45], as well as aiding in the design and planning of living spaces [38, 45]. As Bhandari et al. argued most of their asylum seeker participants accessed the internet using mobile devices at least part of the time [12]. However, due to digital literacy and language barriers, Jensen et al. revealed that digital tools can both aid and hinder the integration process of refugees [27]. Thus, there are many challenges in designing technologies for RAS (refugees and asylum seekers). Talhouk et al. identified challenges in integrating health technologies for Syrian refugees in Lebanon, highlighting that varying resources, mobility of refugees, and perceptions of refugees' technology and health literacies can affect the success of such initiatives [50]. Lengyel et al. also suggested that online technologies can be a resource for RAS while also exposing them to racism [33]. Literature has increasingly engaged with the issue of refugees' access to public services, focusing on aspects such as user experience, language barriers, data privacy, and accessibility. Although these themes contribute significantly to the existing body of knowledge on refugees in CSCW & HCI, there is a gap in understanding the role that mediating service providers and non-governmental organisations play during refugees' resettlement and refugees' experience towards the services facilitated by the service providers. As such, the primary objective of this research is to explore the systematic support that newly-arrived humanitarian entrants received from the service providers.

## 2.2 Early-Stage Settlement Services and Service Providers

The refugee-focused service providers play a pivotal role in facilitating the integration and well-being of refugees by connecting them to essential services such as housing, healthcare, education, and employment [46]. As the definition of mediating organisation proposed by Berger and Neuhaus, "those institutions standing between the individual in his private life and the large institutions of public life", the non-governmental refugee-focused organisations and service providers serve as mediating organisations [11]. These service providers create a sense of welcoming, belonging and appreciation among individuals and endeavour to integrate refugees into the local community [25].

While in the field of CSCW & HCI research, the mediating organisations or service providers to refugees are not widely studied, other research fields have shed light on their effects and existing issues. Üstübici [54] conducted interviews, observations, and informal discussions with service providers in Istanbul, highlighting their crucial role in supporting refugees. This role encompasses unpacking hospitality, facilitating refugee reception, and fostering social cohesion. The author emphasized the importance of focusing not only on interactions between refugees and service providers but also on interactions between service providers and their fellow citizens, which is essential for mitigating public resentment towards refugees [54]. Ekmekcioglu et al. [18] studied the practicality of remote work mode for social workers from humanitarian organizations and concluded that pivoting to remote is not favoured or feasible under a low-resource context.

Jewson et al. [28] studied the experiences and issues of service providers in supporting refugees, in Geelong, Australia. Through semi-structured interviews with 22 participants who worked in refugee-specific roles or other related service providers, they concluded that the barriers negatively influenced the effectiveness of the services provided to refugees in regional and rural areas, which include political elements and media, the locality of resettled refugees, and structural and organisational cultural issues [28]. McIntosh et al. [34], in their New Zealand-based study, adopted the Ketso method with stakeholders from 34 refugee-focused service providers, a facilitated workshop technique, to gain insights on how to better organise the welcome offered by service providers. They found seven common emerging themes identified by participants including relationships to challenge discrimination, education, resources, policy, and service delivery, understanding clients' needs, empowerment, and capacity building, welcome and nurturing, research and advocacy for change [34]. However, it is essential to note that they predominantly collected this data from the viewpoints of service providers, as all participants involved in these studies were either employed by or affiliated with service providers.

In contrast, O'Higgins [39] focuses on the perspectives of young refugees and their experiences with service providers in the UK, through engaging them in focus groups and other group activities. The author argues that social workers tend to deny support if young refugees do not meet the level of vulnerability they were expected to, as social service providers consider agency and vulnerability as two opposing and exclusive categories and discuss the significance of engaging young refugees in participative group work for peer support [39]. While our research also focuses on the newly-arrived humanitarian entrants' experiences with service providers, it covers a more diverse range of ages of participants and proposes ideas on how technologies can be utilised for improvement.

Many have investigated the initial services provided to refugees upon their arrival in different countries. For example, in the UK, newly-arrived refugees are provided with information covering group charters, introduction to the neighbourhood, property instructions, emergency contacts, maps of the local area, public transportation timetable, a group charter, contacts of the service provider, as well as photos with a welcoming message [63]. They also receive initial goods for the first two weeks in the UK, comprising £200-300, pre-paid transportation cards, SIM cards, groceries, toiletries, warm clothes or vouchers to purchase clothes if refugees arrive in winter, and a laptop or tablet to access information online [63]. In addition, under the national context of New Zealand, the organisation Save the Children, designed a dedicated kit for refugee children to facilitate their school transition [64]. Different from the adults, the kit for refugee children contains a welcome letter or video, photos or a tour of a New Zealand school, instructions for preparing for school, a school bag, uniforms, tips for making friends, an introduction to New Zealand culture, and Māori greetings and phrases [64]. In the U.S., the ongoing services lasting for 3~12 months to support refugees' integration contain English training, case management, and employment support [35].

Social Inclusion Theory emphasizes the importance of enabling all individuals to participate fully in society [55]. For refugees, social inclusion involves overcoming barriers related to language, culture, employment, and access to services [8, 57]. This theory provides a lens to examine how resettlement services can support refugees in becoming integrated members of their new communities. Recognizing the known difficulties refugees encounter in accessing public services, prior research in HCI has dedicated significant efforts to enhance the accessibility of resettlement-associated services. Dahya et al. explore the intersection of technology and social inclusion for refugee women in the United States and provide insights into how technology education and access can support refugees' integration and participation in society [16]. Coles-Kemp and Jensen [15] found that even though refugees face uncertainty and insecurity in their daily lives, they focus more on the benefits of accessing services over the vulnerability or the protection from threats in the services. Their research also indicates that health, time, and language-related challenges represent the primary obstacles for refugees when accessing services, rather than issues related to digital literacy or the accessibility of technologies. Similarly, Talhouk et al. [52] found that Syrian refugee women in rural Lebanon commonly utilize mobile phones and the messaging app WhatsApp. In addition, they investigated the potential of ICTs to remotely assist these women in accessing antenatal care services. The communication between refugees and service providers has also been deemed essential for refugees' integration and resettlement. Brown and Grinter [13] created an app Rivrtran to facilitate communication between newly-arrived refugees and their mentors, where the objective was to enhance the mentors' capabilities and streamline communication processes, particularly in cases where mentors and refugees did not share a common language. This was achieved by introducing a voluntary interpreter and implementing a human-in-the-loop interpretation system. Previous studies have often focused on a particular aspect of service accessibility, aiding specific public service concerns [13]. However, there remains a gap in understanding the complete resettlement landscape for refugees, encompassing all services offered by service providers. Therefore, our exploratory research aims to bridge this gap by providing a holistic overview of resettlement services delivered to refugees during their initial resettlement phase, as well as capturing the refugees' experiences with each of these services. Furthermore, our study seeks to identify areas where service providers can improve their support to facilitate the integration and resettlement of refugees more effectively.

## 3 Method

In this section, we outline the research design and methodology employed to achieve the objectives of this study, which aims to understand the experiences of newly-arrived refugees with the services provided by a service provider, identify potential areas for improvement and explore preliminary design opportunities.

### 3.1 Research Context

The authors collaborated with a refugee settlement service provider for migrants and refugees in Australia. This service provider has helped new arrivals in Australia to settle into their new lives in the new country, forging connections within their communities, seeking employment and educational opportunities, developing new skills, and establishing a sense of belonging. Different from other service providers focusing on a specific service, this service provider offers support and goods that are necessary to newly-arrived humanitarian entrants and connects them to all public resources and services available to them, which acts as a mediating organization. The service provider assisted our research project in locating participants and providing bilingual interpretation services for participants as and when needed. In return for taking part in this project, each participant received a $20 grocery gift card, as a token of appreciation.

### 3.2 Participant Selection and Research Design

Our study employed a two-stage qualitative approach to gather an in-depth understanding of participants' perceptions and experiences of their initial settlement journey in Australia. Following ethical approval from our institution, potential participants were handed an information sheet along with a consent form. We sought participation from newly-arrived refugee adults who arrived in Australia between late 2022 and early 2023. Following this process, 12 participants volunteered to take part in this project – we had 3 participants each from Syria, Ethiopia, Afghanistan, and Iran (6 men and 6 women) of varying ages (from 23 years to 62 years).

The demographic characteristics of the participants are presented in Table 1, along with their educational details, household details, ethnicity and language. It is important to note that the household size indicates the number of individuals migrating together as a family unit to Australia and living in a shared space. After receiving signed consent, participants took part in two consecutive stages of this research, each of which is explained below in detail.

Table 1: Participants' Demographics

| Participants | Age Range | Gender | Educational Level | Household Size | Ethnicity & Language |
|---|---|---|---|---|---|
| P1 | 20s | M | Secondary | 1 | Afghan (Dari) |
| P2 | 30s | M | Secondary | 1 | Iranian (Farsi) |
| P3 | 40s | F | Tertiary | 1 | Iranian (Farsi) |
| P4 | 40s | M | Tertiary | 1 | Iranian (Farsi) |
| P5 | 60s | M | Secondary | 6 | Eritrean (Tigrinya) |
| P6 | 30s | F | Secondary | 7 | Eritrean (Tigrinya) |
| P7 | 20s | F | Secondary | 2 | Eritrean (Tigrinya) |
| P8 | 60s | F | Primary | 3 | Afghan (Dari) |
| P9 | 50s | M | Secondary | 6 | Afghan (Dari) |
| P10 | 30s | M | Tertiary | 3 | Syrian (Arabic) |
| P11 | 50s | M | Secondary | 4 | Syrian (Arabic) |
| P12 | 30s | F | Secondary | 5 | Syrian (Arabic) |

*3.2.1 Stage One.* Stage one comprised a photo diary task assigned to participants, which required them to document their living spaces and utilisation of goods provided by the service provider upon their arrival in Australia, over the course of 5-7 days. Diary studies are a well-established method used for capturing a wide range of information [37]. Typically, this method involves participants recording their experiences throughout the day using written descriptions, audio recordings, and pictures [2, 41]. Diary studies offer significant advantages by capturing data from the participant's perspective, using their preferred communication methods, and within real-world contexts, which is particularly effective for collecting subtle insights and nuanced details, especially for infrequent, sensitive, or hard-to-observe events [43]. Participants were asked to record responses to three major questions and photo shooting tasks every day for 5-7 days: 1) *Item Usage*: Choose an item from the basic essential package provided by the service provider, and record and discuss how it was used and incorporated into their daily lives, capturing photo evidence of its usage; 2) *Perception of Services*: Discuss a service provided by the service provider, how it facilitates the re-settlement or integration, and any suggestions on potential improvements, including taking a photo of the service if participants were receiving it during the photo-diary duration; 3) *Accommodation and Personal Changes*: Record any personal changes made to their current accommodations to meet their specific needs, any issues of their current homes and lives, and future expectations, including taking photos of their changes or modifications to their homes. Participants were empowered to select their preferred method for completing the photo diary, either digitally or via paper, based on their level of digital literacy or preference. Figure 1 shows examples of photo diaries from our participants. The photo diary submissions were made daily. Paper-based photo dairies were translated by interpreters and reviewed immediately before stage two of this research, i.e., the interview phase. Those opting for the digital format submitted their daily photo diary tasks via email, WhatsApp, or direct messaging the first author. The photo diaries completed in languages other than English were translated by interpreters. One elderly participant (P8) faced difficulty in completing the diaries or accurately expressing themselves due to their limited language literacy, so a young family member of hers assisted in completing the task.

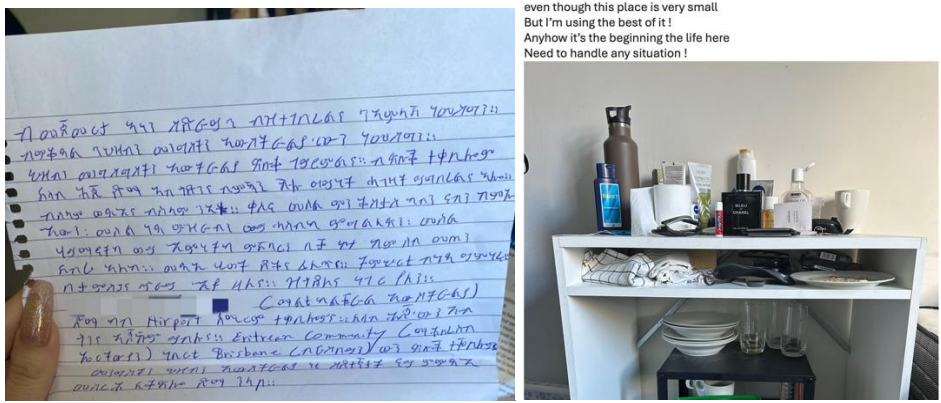

Figure 1. Examples of Photo Diary Collected; 1) Paper-based; 2) Digital, Email-based.

*3.2.2 Stage Two.* Upon completion of stage one, stage two involved semi-structured interviews accompanied by photo-elicitation with participants after they completed the stage-one photo diary. The photos that participants took from the stage one photo diary task served as a conversation catalyst for the interviews in stage two, to inspire rich discussions about their personal experiences, emotions and associations related to their resettlement process and the support they received. Participants were empowered to choose the physical locations of their choice for the interviews. Therefore, the interviews took place in participants' homes or other locations that were convenient for them. During the interviews, participants who were confident in their English proficiency communicated in English, whereas others received assistance from interpreters.

## 3.3 Data Collection & Analysis

Data are collected through the stage one photo diary study over 5 to 7 days, and stage two which comprised of semi-structured interviews, lasting on average of one hour. The former captured visual documentation of participants' living situations and interactions with the basic essential package, while the latter gathered in-depth insights guided by the photo diary. The collected data, including interview transcripts and photo diaries, underwent thorough thematic analysis [14]. This entailed comprehensive readings of the materials, followed by systematic text coding. An inductive coding approach was utilized to initially cluster the services, then pinpoint and categorize the needs of participants that fall into each service category [20]. Besides grouping services based on the timeframe, the initial development of the codebook involved labelling recurring words and phrases, followed by hierarchical grouping of similar codes. Subsequently, the codebook underwent multiple rounds of revision by the second and third authors before being further refined by the first author. Consensus meetings were held to resolve coding discrepancies and improve the reliability of the coding process. Through this process, emerging themes and patterns were identified thereby providing insights into participants' perceptions of the early-stage resettlement and their evaluation of the support received.

## 3.4 Positionality Statement

The authors recognize the importance of acknowledging subjective perspectives and potential biases in studying marginalized and sensitive populations. The second and third authors have worked with refugees for over ten years and have an awareness of their vulnerabilities and potential risks. The team was aware of the potential conflict of interest in collaborating with a service provider whose services were being closely looked at. Our diverse backgrounds and disciplinary perspectives may have influenced our interpretations; however, we have aimed to take the refugee-centric perspective and be truthful to their needs. We are committed to reflexivity, ensuring our research reflects participants' experiences authentically and ethically. In essence, the qualitative research design, inclusive participant selection criteria, and mixed method data collection techniques collectively contributed to a preliminary and multifaceted understanding of the early-stage resettlement experiences of newly-arrived humanitarian entrants. Furthermore, this approach sheds light on the effectiveness of the services and support offered during this crucial phase of their journey from the viewpoint of those who were at the receiving end of the services.

## 4 FINDINGS

This section presents the research findings, derived from an in-depth exploration of the multifaceted early-stage resettlement process for newly-arrived humanitarian entrants, where the authors identified a diverse range of services offered and associated challenges by individuals under the auspices of the local service provider. As depicted in Figure 2, the services are categorised based on the timing and duration of their provision to newly-arrived humanitarian entrants throughout their resettlement journey. There are three key stages of the refugee resettlement process, which include: immediate services upon arrival, initial settlement services, and ongoing services. Each of these stages represents a distinct phase in the resettlement journey, addressing specific needs and challenges encountered by newly-arrived refugees. The immediate services upon arrival focus on urgent needs, such as providing basic resources (e.g., transportation cards, SIM cards), airport pickup, and temporary accommodation arrangements. The initial settlement services cover the initial integration steps, including access to essential public resources, language training, housing, and employment assistance. This stage helps refugees begin their adaptation to local systems and environments. The ongoing services reflect the longer-term needs of refugees, particularly in healthcare, financial stability, and emotional well-being. These services aim to support refugees' long-term independence and successful integration into the host society. Throughout each stage, various themes emerged, highlighting the critical role of service providers and the participants' evolving needs. These themes are discussed within each stage to offer a comprehensive view of refugees' resettlement experiences.

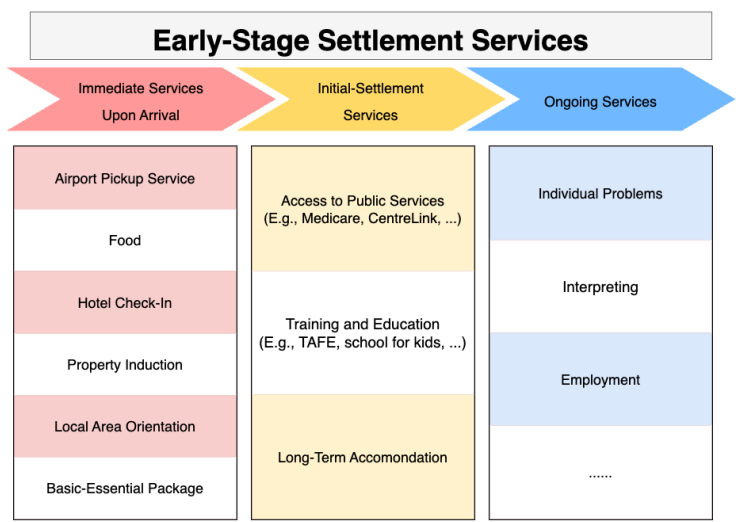

Figure 2: The array of services provided during various stages of resettlement.

## 4.1 Immediate Services upon Arrival

Upon their arrival, newly-arrived humanitarian entrants are greeted by a bilingual case worker, who is tasked to assist with a range of services. These comprehensive services include airport pickup and support at both the arrival and transferring terminals, providing meals, facilitating hotel check-in, and property induction on the very first day upon arrival, along with a basic essential package provided for the first few days in Australia. Orientation to the local area is conducted on the second day. The basic essential package, a curated collection of indispensable items catering to daily necessities, encompasses a mobile phone with a SIM card, vouchers for shopping at supermarkets, and a local transportation card.

### 4.1.1 Sense of Welcome and Reassurance

Hospitality is a core notion embodied in the initial-stage resettlement services provided to newly-arrived humanitarian entrants. Many services create an environment of hospitality and friendliness, helping newcomers feel embraced and develop a sense of reassurance in the new environment. Participants frequently highlighted their positive experiences with the welcoming atmosphere fostered by social workers. For example, P10 initially worried that his family might not be able to find the social worker at the airport. However, he easily identified the social worker holding a large placard at the arrival gate. P10 remarked:

> *"The person is almost, could say waiting for us at the gate of the airport. It was very, very easy to find him. And he guides us through the checkout, step by step. The person was very helpful. And we personally had a problem and he helped us to solve it. We couldn't find our visa and he got us to the person who could help us, and he said wait here is normal procedure and they find our visa and re-enter immediately. It was very helpful."*

Additionally, the orientation service conducted on the second day of each participant's arrival, which included a city tour and introduction to local facilities, further demonstrated the commitment to hospitality, ensuring newcomers felt cared for and supported. P11 expressed his joy about the city tour and emphasised its significance in giving newcomers a sense of welcome. P10 also highlighted this aspect of the orientation, stating:

> *"The second day they immediately met us in the morning, the guy took us for a tour in the city, showing the city council... They introduce the buses, the streets, the lights, things like that. Because everything is new for us, new rules. We don't have it in our country. That's why they introduced us to everything in the first two days."*

Through participants' responses, it is evident that the local area orientation service not only provided newcomers with an enjoyable experience of exploring the city, but also delivered feelings of hospitality, friendliness, and a sense of being embraced and accepted in their new home country and city. It familiarized them with the overall settings of the local area, local culture, and everyday practices, enabling them to integrate more effectively and quickly into the new country. One identified future improvement of the orientation service was the staffing level. In Participant P2's case, the social worker was assigned to two families simultaneously, resulting in limited time for local area orientation after the bank appointments. Consequently, P12's family was unable to receive information about essential information such as how to purchase groceries and use vouchers. Fortunately, P12 mentioned that a fellow hotel resident who spoke the same language assisted. This situation underscores the need for adequate staffing levels to ensure that all newcomers receive comprehensive orientation and support.

### 4.1.2 Tangible Support for Physiological needs

The initial-stage resettlement services effectively fulfil the basic physiological needs of newly-arrived humanitarian entrants, ensuring they have the essential resources required for daily living, such as food and shelter. It was learnt that newly-arrived humanitarian entrants were offered meals upon arrival, catering to their immediate nutritional requirements. Hot food was provided for participants who arrived during mealtime; otherwise, the service provider would offer bread and fruit. For example, P10 was served a chicken meal, whereas P4 received bread with jam and orange juice. Most participants expressed gratitude for the efforts taken - for example, P11 noted, *"The first meal when we arrived, it was so helpful, so help, because we don't know where to find the restaurants".* The service provider accommodates newcomers at hotels before finding long-term housing solutions for them. Regarding the hotel check-in services, all participants reported that the process was quick. The social worker took them to the hotel immediately after picking them up from the airport, allowing them to rest on the first day.

The provision of the basic essential package played a crucial role in meeting the basic needs of newcomers by providing transportation, access to information, technologies, and everyday goods. Many participants highlighted the importance of the transportation card in the package for navigating their new environment, and some participants still kept and continued using the transportation cards provided by the service provider till now. As P10 noted,

> *"I'm still having the transportation card, you wanna see? Two transportation cards, one for me, one for my wife......I think the transportation card is the most useful one (among all the items in the basic essential package)......The card helps you, especially at the beginning, when they learn how to use the public transportation, it helps a lot."*

Another highly regarded item was the SIM card. Several participants emphasized how this SIM card facilitated their lives. For instance, P4 highlighted the significance of staying connected, while P12 emphasized the role of the SIM card in overcoming weak internet signals during their hotel stay. Individuals or families with smaller households expressed contentment with the voucher's value, finding it adequate for their needs. P3 could not remember the exact number of vouchers, but she told us there were two or three cards of $50, and it was sufficient for her to buy food and groceries, as the service provider had already provided personal hygiene products such as shampoo, toothpaste, and other essential hygiene products. While some families with bigger households mentioned they received $350 vouchers for the first two weeks, P8 and P5 noted that they were able to request more vouchers if it was insufficient. For example, in P5's case, as Centrelink had a delay in processing their payments, so they asked for an extra $200 voucher,

> "So when they came they give us, as I told you, the voucher. It was like $350. And also, we didn't have Centrelink. Obviously, they got everything sorted like the bank and everything. But the Centrelink was delayed for a month and two weeks. We have to just keep waiting for the social worker to send us voucher again. So she will send like $200 voucher for the Woolworths for us to go and get food or whatever."

These comprehensive efforts underscore the resettlement services' dedication to ensuring that newly-arrived humanitarian entrants have their basic and physiological needs met, facilitating a smoother transition and integration into their new environment.

### 4.1.3 Tailoring Support to Unique Needs

In addition, it is worth mentioning that besides conventional transportation and lodging provisions, another important aspect of the resettlement services is tailoring support to address individuals' unique needs. Each newcomer or newly-arrived family may have different physical or mental conditions, especially families with disabled members. For example, P12 shared a touching narrative of how the service provider's dedicated support made a profound difference in her family's transition journey. P12's family included members with severe disabilities, which required specialized attention. P12 mentioned that her family informed the service provider of their situation on short notice, yet the provider demonstrated exceptional commitment by arranging additional staff support in response to their needs. While the other family members were brought to the hotel, the disabled family members were taken to the hospital with medical assistance. Such personalized efforts were deeply appreciated by P12, who expressed her gratitude for the crucial role the service provider played during this critical phase of their lives. However, the hotel room provided to P12's family wasn't suitable for wheelchairs, due to the narrow doorways and stairs in the room. This example embodies a potential improvement for service providers.

Another area for improvement is the provision of phones and vouchers. While some participants were satisfied with the mobile phones included in the essential package (Figure 3), others suggested that the quality and functionality of the phones could be improved. For example, P7 noted issues with the phone freezing and not working well: *"The phone didn't really work. So, we had a problem with a phone… it froze. It freezes, like if someone calls you, it just freezes, and you can't pick it up or like something."* These concerns about device quality were common, especially since many participants had brought their own phones to Australia, rendering the provided devices redundant. In light of this feedback, P4 offered a constructive suggestion to improve the provision of mobile phones. He proposed that the service provider could give newcomers the option to choose between receiving a mobile phone or an equivalent monetary voucher. This approach would allow participants to tailor their choices based on their specific circumstances and preferences. P4 also emphasized the significance of being allocated a 'memorable' phone number to facilitate integration, highlighting the potential benefits of providing flexibility in phone selection and associated use within the chaos of initial settlement. Similar advice was also proposed by P5 regarding the vouchers, who shared a story about his children playing in the park and wanting to buy ice cream like the other kids who were enjoying treats. Nevertheless, he was unable to purchase the ice creams for them because the vouchers could only be used at the major supermarket chains. Hence, he proposed an improvement idea: providing monetary support to families in special situations. Overall, these examples suggest that it is important for service providers to empower newcomers to choose the support or services that best meet their needs, ensuring a more personalized and effective resettlement experience.

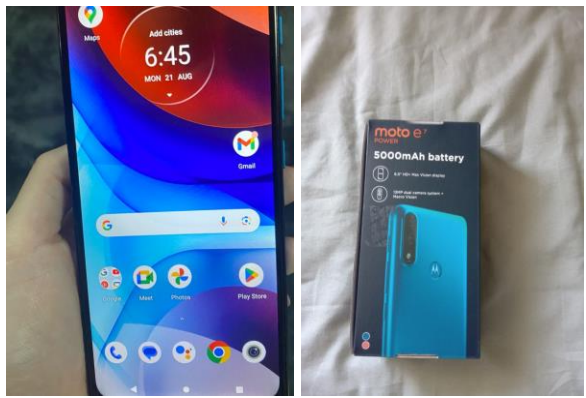

Figure 3: The mobile phone provided by the organization.

### 4.2 Initial Settlement Services

Once the newly-arrived humanitarian entrants settle in the hotel and receive immediate services upon arrival, the service provider continues to assist with initial settlement services. These services include helping them access local services and resources. The newcomers also receive training and education arranged by the service provider. Since the hotel is temporary accommodation, the service provider also assists newly-arrived humanitarian entrants in finding long-term housing. We identify and unfold the insights of participants on initial settlement services based on three perspectives as below.

#### 4.2.1 Infrastructural and Social Adaptation.

To support the adaptation of newly-arrived humanitarian entrants to the local systems and environment, the service provider arranges for them to connect with essential services provided by Services Australia, an Australian Federal Government department. These government services include *Centrelink*, which provides income support and other payments, as well as *Medicare*, the health welfare program that offers Australian citizens and permanent residents access to healthcare. Additionally, the process includes opening bank accounts and completing other necessary steps for preparation. For instance, P12 noted that the service provider provided an interpreter to assist her with the bank appointment and all the paperwork needed.

The service provider also held its own in-house training sessions for all newly-arrived humanitarian entrants to give them a comprehensive understanding of the local systems and environments. These sessions covered life skill topics and five major modules: transport, health, education, community participation, law, and Centrelink. The overall aim of these sessions was to provide an overview of the Australian systems and environments. These sessions were highly praised by all participants, and the information they acquired was found very beneficial for their daily lives thereafter. For example, the service provider not only introduced basic information about Australian systems but also introduced Australia's culture and how to integrate into the local communities. As P9 told us,

> "They told us how you contact with police when you need, how you contact with ambulance, triple zero, and how you contact emergency, contact number, and when you need an interpreter, what's the number you're calling, and everything. And that was very helpful for helpful… So the program was very helpful. They taught us about the Australian government program, and how you deal with your neighbours, the people, like with your family, like everything in Australian culture."

P10 informed us that the service provider provided interpreters for the sessions, catering to all different languages for newly-arrived humanitarian entrants. However, he proposed a piece of advice regarding the timing and duration of these sessions. He mentioned that currently, the sessions ran from 8 am to 2 pm, with a one-hour break. There was a possibility of people becoming distracted and fatigued, particularly when listening to unfamiliar languages while awaiting interpretation. He recommended breaking down the sessions into shorter durations and spanning them over two or three days.

Participants also highlighted the critical need for social connectedness as an essential step for their integration into the Australian community. P10, who had relatives in Australia, informed us that in their culture, people tended to prioritize recommendations from friends or family due to their cultural background. This reflects a common resettlement challenge among refugees and their mistrust towards local service providers or the government. As a result, P10 initially attempted to address banking and Centrelink matters independently after seeking advice from their friends but encountered issues and subsequently sought assistance from the organization.

> " When we make our account, like, my wife, I connected for her (to) Centrelink, I need to get connected too, but my own, there was a problem. I couldn't connect. Then they (the service provider) came and helped us with the matter. Also, they help us to connect it later."

Participant P9 emphasized the importance and need for connecting with people from the same cultural background as a crucial foundation for social integration. P9 shared that he and his family lacked social connections in Australia. He expressed their desire to connect with individuals or families who spoke the same language and shared similar resettlement experiences, especially given their ongoing process of learning English. Their neighbors were all native Australians, and the lack of English language proficiency was proving to be a communication barrier. P9 suggested that if the organization could arrange a service aimed at enhancing social connectedness among newly-arrived humanitarian entrants and those who have previously resettled and share a similar cultural background, it would be greatly appreciated.

*4.2.2   Individual Development.*

The service provider facilitated the individual development of newly-arrived humanitarian entrants by helping them register for English courses at Technical and Further Education (TAFE), Australia's largest vocational education and training provider. P3 expressed immense gratitude for the English training and mentioned that the financial support provided by the Australian government was sufficient to cover her living expenses even without employment, allowing her the wonderful opportunity to dedicate herself to full-time English learning for an entire year.

Additionally, the service provider assisted with enrolling children into school. For P10, who had a young child, the service provider arranged day-care services. Participant P8 highlighted that the organization tailored its approach to each child's situation. For her younger child with disabilities, the organization ensured enrolment in a special school. Participant P5 even expressed that he felt a renewed sense of hope for his family's future when he was asked about his outlook on their future life in Australia during the interview because his children were now able to receive education,

> *"So also, because the school here for the kids, they can study, they have a bright future, in the future, finishes school, look after themselves. So that's like, for us, that's a big thing. So, we can see a bright future for the kids, or either for us, the family."*

Through these insights from participants, it is evident that the training sessions and education provided by the service provider were extremely beneficial for all refugees. These services not only accelerated their integration into the Australian environment and helped them overcome language barriers, but also brought hope to their lives and instilled in them a positive vision for their future in Australia.

*4.2.3 Sense of Home and Belonging.*

Creating a sense of home and belonging is critical for the settlement of newly-arrived entrants. According to P12, she nourished a sense of home after residing in Australia, that she had never experienced before,

> *"We have been asking for asylum here in Australia since 2016, we waited for a long time. We didn't have a home since 2016, and until we moved here. That's like home, that's home for us. Even when we go to school now (learning English at TAFE), we are just waiting to come back to our home and close the door and we feel at home here."*

P12 and her family were arranged a long-term accommodation by the service provider. As mentioned, since they had family members with disabilities, it would be inconvenient to live in an accommodation with stairs. The service provider made a special effort to identify and recommend one stair-free house. P12 and her family were very happy with the house that the service provider provided to them, which had been recently refurbished and featured adequate rooms and space. They also saved money to purchase new furniture and decorations for their home, including a large TV for the entire family to enjoy and two additional large sofas for the disabled family members to lie down. They regarded these practices as a way of establishing their new home in the host country, and found joy in the process, accompanied by a sense of accomplishment, and belonging.

The authors were informed that newly-arrived humanitarian entrants are offered a maximum of a three-month free stay at the designated hotel and then transition to long-term accommodation after they connect to Centrelink. These newcomers had two options in seeking accommodation:

1) The first option was for the newcomers to search for rental properties on their own and apply for rental financial support from the service provider. Once the application is permitted, they can proceed to move in;

2) The second option was to allow the service provider to arrange rental properties for them. While the social worker would provide basic information on the available properties, the newcomers would be unable to inspect the property physically. If the newcomers decided to move in, they must sign a lease for at least six months, which was binding.

Since the rental period is at least six months, unlike the hotel where they initially stayed for only a few days, newcomers need more than just a place to sleep; they need a place to live, a place that feels like home. Therefore, the participants expressed their needs for a 'home' and their expectations that the rented house could meet their needs.

Participants who were able to rent properties by themselves were mostly satisfied with their accommodation since they could physically inspect and select the properties. An example of this is P10, who rented property with the help of his relatives,

> "This is the most difficult process, finding a house and applying for it. We have some relatives here; they know the owner of this house. So, we talked to the owner. And he said, okay if you want the house, no problem, we'll go through the paperwork. And when it came to the bond, they applied for a bond loan for us. The service provider applied for us, and we got a fast approval, saying they got the approval. Wow. That was very fast. And it was very, very helpful."

P11's family also opted to independently secure their current residence, as they preferred not to consider the second option due to its inspection restrictions, so they applied for rental financial support to pay for the bond. They told us they could afford their current rent and reported being very satisfied with their living conditions, as shown in Figure 4. Their current residence was a 2-storey house, with a large living area, kitchen, and a car park on the lower level, and three well-maintained bedrooms on the upper level. While the two children occupied a smaller bedroom each, the parents resided in the master room. They were all content with the overall size of the house, as well as the individual rooms.

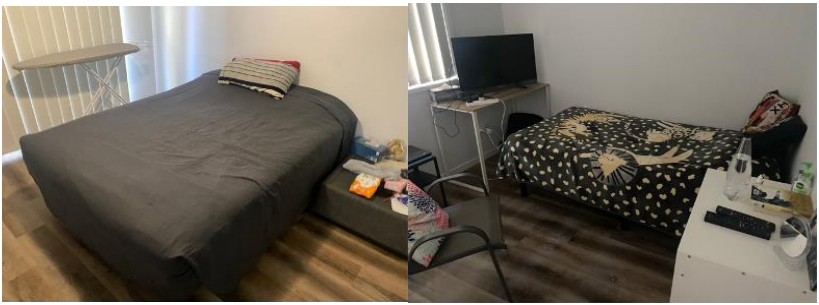

Figure 4: Bedrooms of P11's current long-term accommodation: a) master room for parents; b) single room for children

While other participants lacked personal connections in Australia, they followed the service provider's accommodation arrangements. It was found that families with bigger households tended to be mostly pleased with the provided accommodation, as they have the support of their family members and their primary need for a home is to stay together with their loved ones. For example, Participant P8 expressed that her need for a home is fulfilled as long as she is staying with her two sons. Participant P9 also expressed his gratitude for the house he lives in with his family, emphasizing that it is a safe place for them. The support and presence of family members significantly contribute to their satisfaction with the accommodation. However, the leasing situation and the no-inspection policy led to certain challenges for individuals settling in Australia alone. Some participants who came to Australia alone expressed that the long-term accommodation has not met their needs for a home. Living in shared housing arrangements can be particularly challenging, as individuals need to communicate with their roommates and occasionally make compromises regarding their own needs. For instance, P3, who lived in a two-floor shared housing, mentioned that there was a minor issue with her flatmate who was sharing the bathroom with her. She told us that the cleanliness of the house is the most important to her among all other living conditions. However, her flatmate constantly left the bathroom uncleaned after use. Though she tried her best to adapt, she still found it challenging to cope with the situation and remained uncomfortable due to the hygiene conditions.

A family with smaller households also encountered a similar issue, involving conflicts with the neighbors. Participant P7, a young female who came to Australia with her little sister, was assigned to a residence that she thought was located in an unsafe area. From her sociocultural standpoint, she was concerned because she felt that the particular neighborhood where she lived was frequented by intoxicated people, who did not welcome the newcomers and kept disrupting the peace. She shared,

> "The neighbors were always drinking, drinking, drinking. So, when they get drunk, and like around 4 a.m., or whatever, they will throw the glass bottles on the house… And then we complained about it to the office and like to the social worker, they couldn't do anything."

Participants suggest that there is a need for enhancements in the procedures for assessing situations and revising policies, such as reconsidering inspection restrictions or adjusting leasing limitations for unique scenarios. This is particularly crucial for more vulnerable individuals, guaranteeing the well-being of recently arrived humanitarian entrants and ensuring that their needs for a home are met.

### 4.3 Ongoing Services

*4.3.1 Care and Empathy.*

After transitioning to their long-term accommodations, newly-arrived humanitarian entrants may continue to face individual issues, including booking appointments with a general practitioner (GP), handling emergencies, or overcoming language barriers. During this phase of their settlement, the social worker assigned to each family or individual will be responsible for assisting with these problems, and the newcomers heavily rely on the support from their assigned social workers. Many participants expressed appreciation for the efforts of their assigned social workers, emphasising their helpfulness and strong willingness to assist. Participant P3 specifically mentioned that her social worker was exceptionally warm-hearted. She passionately said that the help and support from her social worker were more significant than any material support. In a particular instance, she was dealing with dental health problems and sought help from her social worker to assist her in scheduling an appointment with a dentist due to language and cultural barriers,

> "She (refers to the social worker) followed up with a phone call, and we went to the clinic to do some blood tests and visit GP and they check everything and I'm, you know, satisfied about the medical services. And about that same dentist, I have an appointment, but in the next two months for the dentist, and she (the social worker) followed up to organise this."

It can be seen that P3's social worker's support extended beyond P3's initial request of booking the first appointment, but also cared for and assisted with the ongoing appointments. This level of commitment and personalised care from the social worker had an immensely positive impact on the well-being and successful integration. However, it came to the authors' attention that a few social workers may not demonstrate sufficient empathy as expected. P1 expressed dissatisfaction with his social worker, who always ignored his phone calls or messages. He told us even though the social worker acknowledged his requests or needs, there was still no tangible assistance provided. This point was further discussed by P4, who highlighted the differences in the sense of responsibility among social workers,

> "The first thing is about the social workers, they are different by case by case. For me, that social worker is like a friend, really responsible, and tries to provide me with everything I need in the short term, you know, very soon. But about the other friends of mine, it's not true, because for example, the, maybe, the social worker doesn't answer the phone."

P4 shared that when he asked the organization to schedule a GP appointment, a social worker arrived without a vehicle to demonstrate how to take public transport to reach the clinic, which he already learned during the orientation. During his subsequent appointments, the organization repeatedly inquired if he required assistance in going to the clinic. This duplicated service was unnecessary for him. In contrast, while P4 received redundant services, P1 encountered a critical health situation and contacted his social worker for immediate assistance. Yet the social worker only told him to call the emergency himself, without offering any other support. Therefore, it is important that social workers consistently offer compassionate and effective assistance, to guarantee optimal outcomes for these newcomers. The service provider and stakeholders could formulate policies or assessments to evaluate the support provided by social workers to newly-arrived humanitarian entrants under their care, thereby ensuring the quality and consistency of the services.

### 4.3.1 Building Independence.

Most participants demonstrated a strong desire to work and become self-sufficient, rather than relying on government pensions. The service provider offers guidance in applying for work and some working opportunities. Participants who had a higher level of English proficiency found it easier to find jobs than those who were still learning English. For example, P1, who had basic English language proficiency, had been able to find work doing gardening. Likewise, according to P6, her family members with sufficient English proficiency were able to secure employment soon after resettlement,

> "So my husband and my brother-in-law, because they're good at English, he used to do translating in Sudan, and he started working already......So it's good if you know English, you're good. You can start your job like my husband."

Yet, for participants with a lower level of English proficiency, it was challenging to secure employment. P3 had made a personal decision to go to Sydney in search of a job, but she faced difficulties and was unable to find employment, ultimately returning to Brisbane. Upon her return, her social worker assisted her in finding a new place to live. Participants with a higher level of educational background tended to discover that the job opportunities provided by the organization did not meet their expectations. According to P10, who was currently working as a part-time delivery driver, his previous occupation before coming to Australia was a chemical engineer, and used to work for a German healthcare company as a production manager,

> "I'm doing delivery work right now, but when they say that you have to work as a casual, in one place, picker-packer. It's not the job itself. The job is good. I'm not saying that I am not satisfied. That's not my skills. My skills could be more capable, I'm giving you more things. That's why I couldn't accept it. They provide a lot of jobs. But for me, I couldn't accept it."

P10 sought advice from his relatives who resided in Australia for a longer time and realized if he wanted to look for a desired job, he needed to have a recognized degree and qualification assessment. Therefore, he was doing a diploma in IT with the hope of securing better employment in the future.

## 5 DISCUSSION

This study discussed lived experiences of newly arrived humanitarian entrants' settlement journey. It draws on participant narratives to establish a holistic understanding of the services offered by the mediating service provider. While current literature on settlement services for refugees in CSCW or CHI tends to focus on specific services or support, for example, food [51, 53], health [50, 53], education [33], digital tools and technologies [12, 18, 42], and so on, our research draws on participant narratives and sheds light on the entirety of the services offered by mediating service providers. The research provides a comprehensive understanding of the critical role of mediating service providers in connecting refugees to a broad range of essential services. In our study, we categorize resettlement services into immediate, initial, and ongoing phases. This breakdown aligns with the Social Inclusion Theory's focus on continuous and progressive support for refugees, ensuring their integration into the host society [55]. Moreover, in connecting our findings to the work by Dahya et al. [16], which proposes the framework of refugee women's everyday socio-technical ecosystems, where they categorize refugee women's interaction with technologies by the context, i.e., individual, social, or formal learning. Our framework for categorizing settlement services is distinct in its focus on a timely, progressive order, systematically addressing the evolving needs of refugees from their immediate arrival to long-term integration. This structured approach delineates services into three critical phases. Such categorization provides a structured approach that can contribute to HCI and CSCW research by offering clear stages for intervention and technological support. We discuss some of these implications below.

### 5.1 Design Implications

#### 5.1.1 Understanding Individual Needs.

Based on our findings, service providers and organizations could prioritize and implement measures to enhance several critical aspects of their services. Firstly, service providers could aim to understand the individual needs of newcomers better and accordingly provide them with greater freedom in choosing the right and tailored services. For instance, P12, whose family members were disabled, was not provided with accessible hotel accommodation. The organization could try to understand the needs of newcomers better and communicate more effectively to avoid such problems. As McIntosh et al. [34] stated, that understanding the needs of refugees is a vital aspect of effectively welcoming and aiding in their resettlement. Consistent with their viewpoint, our research also demonstrated the importance of service providers being attentive to individual needs. For instance, P4, who was already familiar with the route to the clinic, was repeatedly asked if he needed assistance in getting to the clinic. This results in a situation where, while some newcomers did not receive sufficient support, P4 received redundant assistance. The organization should implement resource allocation strategies and enhance internal communication among staff to avoid duplication of services and ensure a more efficient and effective service delivery process. In line with Jewson et al. [28], the lack of service coordination and consultation with refugees are the barriers that negatively influence the effectiveness of services, service providers should adapt to refugees' needs, rather than expecting refugees to conform to their service delivery methods. The need for tailored services is further supported by Almohamed et al., who emphasize the importance of addressing the heterogeneity of refugees and asylum seekers (RAS) [4]. It suggests that understanding the various challenges faced by RAS, including their cultural, social, and displacement-related stressors, is crucial for designing effective support systems and technologies tailored to their individual needs [4]. The insights around the provision of mobile phones reinforce this point, where our participants recommended offering more options, such as allowing them to select their phones in-store or, for those who do not require a phone, providing equivalent voucher amounts. Hence, organizations and service providers could investigate adopting a more personalized approach, such as utilizing tools like checklists to gain insights into individual requirements, to avoid unnecessary or duplicative services, ultimately saving both time and resources, which could then be directed towards assisting others with more pressing needs.

To address this, a more holistic approach to service navigation is required. Previous research suggested ICTs are commonly used technology among refugees, and they heavily rely on mobile communication such as Whatsapp [52]. Bhandari et al. also indicated that many asylum seekers were actively using the internet and various technologies to meet a range of needs, including accessing information about public benefits [12]. According to our research findings, newly-arrived humanitarian entrants in the context of Australia are provided with mobile phones and widely use them – ensuring accessibility to ICT technologies. Similarly, Talhouk et al. propose that to lessen food insecurity among refugees, technologies can be used to facilitate better navigation about aid services and enable refugees to query and provide feedback on aid received [53]. Hence, based on the accessibility of ICT technologies among these newcomers, we propose the potential of utilizing ICT technologies for more conducive navigation around the services they would receive. With a clear picture of the overall service flow, the newcomers could identify their individual needs in advance and notify the service providers. This application could also include a digital checklist or questionnaire to capture the specific needs and preferences of newcomers upon arrival, ensuring that services are tailored accordingly. Instead of expecting refugees to request disability-friendly accommodations, service providers should proactively gather information about accessibility needs through initial assessments and ongoing communication with newcomers. With this kind of application, the service provider can identify refugees who have family members with disabilities, like P12, and proactively prepare for their need for disability-friendly accommodations and services upfront. Likewise, service providers can allocate resources more efficiently based on the actual demand and preferences of the individuals, reducing redundancy and waste. For example, for refugees who have their own mobile phones upon arrival, the service providers could save the cost and time of purchasing a new phone.

*5.1.2 Fostering Empathy and Eliminating Discrimination.*

Additionally, service providers could foster greater empathy among both social workers and the local community towards newly-arrived humanitarian entrants. Previous research has indicated that refugees often perceive discrimination from the local community, primarily due to negative stereotypes and the influence of political elements on social media [4, 28]. Our study similarly revealed inhospitable conditions in certain neighbourhoods, for instance, P7, who resided with her sister, experienced incidents like drunk neighbours throwing bottles at their walls during the night-time, while the social worker assigned to them failed to provide any assistance. As Üstübici [54] states the pivotal role of service providers in bridging the gap between citizens and refugees, advocating for empathy emerges as a crucial strategy, he also suggested social contact may be beneficial for mutual understanding between native residents and refugees. While service providers educate new arrivals about Australian culture, there is also a need for the local population to learn more about the refugee journey and their culture, and for service providers to cultivate understanding and empathy within the local communities towards refugees. Viewing refugees as a burden impacts not only the interactions between refugees and host country citizens but also their interactions with service providers [49]. Talhouk et al. [52] revealed an instance in which a Syrian refugee woman encountered a negative attitude from healthcare service providers, where a nurse implied that Syrians have too many children. In our study, we did not encounter any cases where social workers nor the organization showed discrimination against newly-arrived humanitarian entrants, however, it might be worthwhile to ensure the wellbeing of social workers who in many cases are former refugees themselves and may have workload issues and other general life worries.

*5.1.3 Facilitating Social Connections among Ethnic Groups.*

Thirdly, there could be attempts to facilitate social connections among refugees sharing common cultural backgrounds. Mistrust is another commonly shared issue of refugees towards the services or authorities in the host countries [4, 7]. In our study, while not many participants displayed mistrust toward the service provider, they preferred to place their trust in their relatives and friends who resided in Australia for a longer time, or individuals who shared their language or cultural background, over the service providers. This discovery aligns with Almohamed et al.'s [4] findings, highlighting refugees tend to avoid unreliable sources and seek information from individuals whose identities and personal information they know, and proposing to address this information provision gap through peer-to-peer knowledge exchange. As such, they suggest governments to support social events for refugees and their ethnic groups [4]. This point is further supported by the case of P10 in our study, who sought long-term accommodation with the assistance of relatives, instead of relying on the organization's arrangements. O'Higgins' [39] study indicated that the participation and engagement of young refugees can be enhanced through participatory group work, primarily due to their strong emphasis on peer support and forming friendships. In light of this, participant P9 suggested the idea of creating a session where newly-arrived humanitarian entrants and individuals with prior resettlement experience who share similar cultural backgrounds could come together to know one another and build social connections. This incident embodies the significance of establishing mentor-mentee relationships between newly-arrived humanitarian entrants and those who have successfully settled and share a common language. This is particularly crucial for those newcomers with limited proficiency in English. As highlighted by Fozdar and Hartley [24], humanitarian entrants exhibit notably higher levels of participation and engagement within their religious, cultural, and ethnic communities, and provide substantial support to one another, encompassing lending household items, offering transportation, aiding with shopping, providing meals, childcare assistance, and even offering financial loans. Hence, organizations and service providers could investigate the possibility of inviting previously resettled humanitarian entrants to participate in the orientation of newly-arrived humanitarian entrants, allowing them to share their own resettlement experiences and help the newcomers.

Technologies can play a crucial role in facilitating social connections between newly-arrived humanitarian entrants and those who have already settled in the host country, particularly when they share similar ethnic or cultural backgrounds. As Talhouk et al. pointed out, sharing resources within households and communities is a common practice among refugees to cope with food insecurity, and emphasized the need for information-sharing platforms that can support peer-to-peer exchange of experiences and knowledge, enhancing the community's overall resilience [53]. In our study, many participants mentioned the need to interact with previously settled humanitarian entrants or connect with someone from the same cultural background. However, they often lack access to platforms that would enable them to meet such individuals. Given the limited English proficiency among many newcomers, which restricts their access to mainstream social media platforms, there is an opportunity to create a dedicated platform aimed at assisting non-English speakers in connecting with their peers, storytelling and sharing experiences. This platform could feature a multilingual interface, community forums, and mentorship programs where newly-arrived humanitarian entrants can receive guidance and support from those who have successfully settled. As stated by Weibert et al., utilizing digital technologies, especially digital storytelling, provides platforms for personal expression, reflection, and building a digital biographical and collective memory [56]. By fostering these connections, the platform would help build social capital, facilitate knowledge exchange, and provide emotional support, thereby contributing to the holistic well-being and resilience of the newcomers. Additionally, the platform could host virtual events, cultural exchange programs, and peer support groups to further enhance community building and integration.

### 5.1.4 Nourishing a Sense of Home.

Lastly, the leasing issue of long-term rental accommodation requires more attention, as argued by Sampson et al. [44], housing and a sense of place are crucial for fostering a sense of belonging, especially for refugees who have been denied access to a stable home of origin. Previous studies have suggested that suitable housing offers refugees a starting point for re-establishing a sense of 'home' [17, 24]. Consistent with this statement, we also discovered that participants derived enjoyment from the process of altering and decorating their living spaces. For instance, P12 considered home decoration practices as a means of establishing a sense of home and belonging in the host country. This example further reinforces the argument made by Nabil et al. [36], who observed that refugees transformed both public and private spaces within refugee camps through painting and decorating, thus creating a sense of home and fostering a feeling of identity and pride. Yet in our findings, the restriction on inspecting a property before renting or breaking the lease before the 6-month term ends impeded their ability to change and find suitable housing that was more aligned with their liking. Housing policies directly impact the ability of refugees to secure and maintain stable accommodation that meets their needs. Policies that limit the ability to inspect properties or terminate leases early can exacerbate feelings of instability and displacement. As Tachtler et al. stated living circumstances and accommodation can have a significant impact on unaccompanied migrant youth [48]. The lack of a stable, supportive environment hinders the ability of unaccompanied migrant youth to establish routines and feel secure [48]. While it might be impractical for organizations to secure perfect accommodation for newcomers in every aspect, prioritizing and understanding newcomers' housing needs and evaluating the pros and cons of each property well in advance might result in a better match between newcomers and housing options.

We propose a design implication of an online housing database that includes detailed information on available properties, such as the suburb, and number of roommates, and pictures of the accommodation, allowing refugees to select housing that best fits their needs and preferences. Additionally, implementing a flexible leasing policy that allows for property inspections before signing the lease and provides options for shorter-term leases can significantly enhance refugees' housing experiences. Collaborating with local housing authorities and community organizations to secure more suitable housing options and providing financial support for home modifications can further improve housing stability and satisfaction. By addressing these housing challenges, service providers can help refugees establish a sense of home and belonging, which is essential for their successful integration and overall well-being.

*5.1.5 Ongoing Support and Integrated Health Services.*

As the findings presented, many participants had long-term needs for future housing, healthcare, and employment services (e.g., P1, P3, P4, P10). However, they were often limited in addressing these needs individually due to challenges such as language barriers and unfamiliarity with local systems. In line with our findings, Ertl et al. highlighted several barriers refugees face in accessing psychiatric support and healthcare services [19]. These contain individual barriers such as language and cultural differences, and structural barriers including complex bureaucratic processes, lack of coordinated care, and social barriers, e.g., social stigma around mental health and lack of social support networks [19]. In our study, social workers took measures to assist participants with long-term issues. For example, participants P3 and P4 explained that social workers helped them book healthcare appointments and followed up on their upcoming treatments. Yet, P1 and P10 expressed their desire to live independently in Australia, particularly in terms of navigating through systems and achieving financial stability. To ensure sustainability and foster the independence of refugees, we propose that service providers adopt an integrated approach to health and well-being services, ensuring that the physical, mental, and emotional health needs of refugees are addressed comprehensively. This could include the development of a centralized health management system that tracks the medical history, appointments, and ongoing health needs of refugees. By employing this system, refugees can receive timely reminders for medical appointments, access telehealth services, and obtain information on available health resources in their native languages. Such support systems should be designed to foster resilience, enabling refugees to thrive independently while still having access to support when needed.

## 5.2  Limitations

While this study provides insights into the early-stage resettlement experiences of newly-arrived humanitarian entrants in Australia, it has several limitations that should be acknowledged. Firstly, the sample size of 12 participants, though providing rich qualitative data, may not fully capture the diversity of experiences across different refugee populations. Additionally, the study focuses on refugees in Australia, which may limit the generalizability of the findings to other geographical and socio-political contexts. The reliance on self-reported data from photo diaries and interviews could also introduce biases, as participants may present their experiences in a manner influenced by social desirability or recall inaccuracies. Furthermore, the study emphasizes the role of a single mediating service provider, which may not reflect the practices and challenges faced by other organizations or regions. Finally, the dynamic nature of resettlement processes means that the findings represent a snapshot in time and may not account for longer-term adaptation and integration challenges faced by refugees. Future research could benefit from a larger, more diverse sample and longitudinal approaches to better understand the evolving needs and experiences of refugees.

## 6 CONCLUSIONS

In this exploratory study, the researchers delved into the experiences of newly-arrived humanitarian entrants towards the resettlement services provided by non-governmental organizations. The authors conducted both photo diary studies and semi-structured interviews with a cohort of 12 recently arrived humanitarian entrants. Through a comprehensive thematic analysis, we presented an overview of the service framework and presented our findings by categorizing them into three distinct service domains: immediate services upon arrival, initial settlement services, and ongoing services. Drawing upon the feedback received from the participants, we identified the needs of the participants and areas of improvement. Additionally, we put forth design implications to enhance the services provided.

## ACKNOWLEDGMENTS

The first author used ChatGPT for copy-editing the author-generated content of the paper. ChatGPT is not used for generating any new content for this paper.